# Hessian-informed Hamiltonian Monte Carlo for high-dimensional problems

Mina Karimi[1*], and Kaushik Dayal[1], Matteo Pozzi[1]

[1]Department of Civil and Environmental Engineering, Carnegie Mellon University

Please address correspondence to:

Email: minakari@alumni.cmu.edu (Mina Karimi)

# Hessian-informed Hamiltonian Monte Carlo for high-dimensional problems

Mina Karimi[a], Kaushik Dayal[a], and Matteo Pozzi[a]

[a]Civil and Environmental Engineering, Carnegie Mellon University, Pittsburgh, Pennsylvania, USA
E-mail: minakari@andrew.cmu.edu, Kaushik.dayal@cmu.edu, mpozzi@cmu.edu.

ABSTRACT: We investigate the effect of using local and non-local second derivative information on the performance of Hamiltonian Monte Carlo (HMC) sampling methods, for high-dimension non-Gaussian distributions, with application to Bayesian inference and nonlinear inverse problems. The Riemannian Manifold Hamiltonian Monte Carlo (RMHMC) method uses second and third derivative information to improve the performance of the HMC approach. We propose using the local Hessian information at the start of each iteration, instead of re-calculating the higher order derivatives in all sub-steps of the leapfrog updating algorithm. We compare the result of Hessian-informed HMC method using the local and nonlocal Hessian information, in a test bed of a high-dimensional log-normal distribution, related to a problem of inferring soil properties.

## 1 INTRODUCTION

Uncertainty quantification is a challenging and crucial topic in many fields. We consider the problem of quantifying the uncertainty of large-scale inverse problem addressed by the Bayesian inference framework. Many algorithms have been proposed to characterize uncertainty of the inferred solution, from Markov Chain Monte Carlo (MCMC) methods to variational techniques such as parametric mean-field and Stein variational method (Chappell et al., 2008, Liu & Wang, 2016).

Traditional MCMC sampling methods such as Metropolis-Hastings MCMC (MH-MCMC) are powerful methods that can be applied to a wide range of problems. However, this method is inefficient for exploring high-dimensional nonlinear parameter spaces. There have been many efforts to improve the performance of the MH-MCMC method (Herbst, 2010, Petra et al., 2014). Metropolis Adjusted Langevin Algorithm (MALA) is an effective method that uses the local gradient information to speed up the Markov process (Roberts & Stramer, 2002). Recently, Karimi et al, 2021 investigate the use of the second derivative information at the maximum-a-posteriori (MAP) point to enhance the MALA algorithm, which shows a significant improvement in the performance of the method, specifically for exploring non-Gaussian distributions.

Hamiltonian Monte Carlo (HMC) is another family of methods that uses the local gradient information and can explore faster than the traditional MH-MCMC method (Neal, 2011). However, when the method is applied to higher dimensional problems, exploration using the HMC method is more challenging, and efficiency decreases.

Girolami & Calderhead, 2011 proposed the Riemannian Manifold HMC (RMHMC) algorithm, which uses the second derivative local information to increase the convergence speed of exploration. Nonetheless, calculating the local Hessian matrix for high-dimensional problems in many cases is expensive. Bui-Thanh & Girolami, 2014 and proposed using the Fisher information at the MAP point. Karimi et al, 2021 investigated the performance of different Hessian-informed sampling methods for non-Gaussian high-dimensional distributions.

In this paper, we study how to quantify the uncertainty in a high-dimensional Bayesian inverse problem. We investigate the effect of using second derivative local information and Hessian at the MAP point on the performance of a HMC scheme when the posterior distribution is a log-normal high-dimensional probability density.

## 2 Metropolis-Hasting MCMC

MH-MCMC method defines a random walk with samples which are taken from a proposal density $q$. Usually, we consider an isotropic Gaussian proposal density as follows:

$$q(\mathbf{\theta},\mathbf{y}) \propto \exp\left(-\frac{1}{2}\|\Delta t^{-1}(\mathbf{\theta}-\mathbf{y})\|^2\right) \quad (1)$$

where $\mathbf{\theta} \in R^d$ is the current value of the random variable, $\mathbf{y}$ is the new candidate value, and $\Delta t$ is a fixed step size. The acceptance criterion of this algorithm is defined as follows:

$$\alpha = \min\{1, \exp(J(\mathbf{\theta}_k) - J(\mathbf{\theta}_{k+1}) + \Delta q)\} \quad (2)$$

where $\Delta q = \log q(\theta_{k+1},\theta_k) - \log q(\theta_k,\theta_{k+1})$, and $J(\mathbf{\theta}) = -\log(\pi(\mathbf{\theta}))$ is the negative of log of target distribution, $\pi(\mathbf{\theta})$.

## 3 Hamiltonian Monte Carlo Method

In the HMC method, we consider a random parameter space as $\mathbf{\theta} \in R^d$ with density function $\pi(\mathbf{\theta})$ which can be interpreted as location, and an auxiliary variable $\mathbf{p} \in R^d$ with density $N(\mathbf{p}|\mathbf{0},\mathbf{M})$ which can be interpreted as velocity, where $N(\cdot|\mathbf{0},\mathbf{M})$ is a Gaussian distribution with zero mean and covariance matrix $\mathbf{M}$, also called the "mass matrix". The Hamilton function can be defined as the summation of negative of log of these two densities:

$$\mathcal{H}(\mathbf{p},\mathbf{\theta}) = J(\mathbf{\theta}) + K(\mathbf{p},\mathbf{\theta}) \quad (3)$$

Where potential function $J(\mathbf{\theta})$ is defined above, and

$$K(\mathbf{p},\mathbf{\theta}) = \frac{1}{2}\mathbf{p}^T\mathbf{M}^{-1}\mathbf{p} + \frac{1}{2}\log\{|\mathbf{M}|\} + \text{const}$$

is the kinetic energy.

Usually, for low-dimensional and linear problems, the mass matrix is assumed constant ($\mathbf{M} = \beta\mathbf{I}$).

To update the variables $\mathbf{\theta}$ and $\mathbf{p}$, the Hamiltonian equations are used to find the derivative w.r.t time $t$. By considering constant mass matrix the Hamiltonian equations are:

$$\frac{d\mathbf{\theta}}{dt} = \frac{\partial \mathcal{H}}{\partial \mathbf{p}} = \mathbf{M}^{-1}\mathbf{p}$$
$$\frac{d\mathbf{p}}{dt} = -\frac{\partial \mathcal{H}}{\partial \mathbf{\theta}} = -\frac{\partial J(\mathbf{\theta})}{\partial \mathbf{\theta}} \quad (4)$$

To implement the updating step, we use the leapfrog algorithm (Girolami & Calderhead, 2011):

$$\mathbf{p}_{k+1/2} = \mathbf{p}_k - \frac{\Delta t}{2}\frac{\partial J}{\partial \mathbf{\theta}_k}$$
$$\mathbf{\theta}_{k+1} = \mathbf{\theta}_k + \Delta t \mathbf{M}^{-1}\mathbf{p}_{k+1/2} \quad (5)$$
$$\mathbf{p}_{k+1} = \mathbf{p}_{k+1/2} - \frac{\Delta t}{2}\frac{\partial J}{\partial \mathbf{\theta}_{k+1}}$$

Moreover, the acceptance criterion of the HMC method is defined as follows:

$$\alpha = \min\{1, \exp(\mathcal{H}(\mathbf{p}_k,\mathbf{\theta}_k) - \mathcal{H}(\mathbf{p}_{k+1},\mathbf{\theta}_{k+1}))\} \quad (6)$$

There have been many efforts to improve the performance of the HMC method for non-Gaussian high-dimensional distributions (Girolami et al., 2009, Karimi et al., 2021, Lee & Vempala, 2018, Chen et al., 2022). The choice of mass matrix is one of the factors that can affect the performance of HMC method. Below we describe and compare some of these algorithms.

### 3.1 Riemannian manifold HMC method

Riemannian manifold HMC (RMHMC) suggests the probability density of $N(\mathbf{p}|\mathbf{0},\mathbf{G}(\mathbf{\theta}))$ for the auxiliary vector $\mathbf{p}$, where $\mathbf{G}(\mathbf{\theta})$ is defined as the second order derivative of the potential function $J(\mathbf{\theta})$, ($\mathbf{G}(\mathbf{\theta}) = \nabla_\theta^2 J(\mathbf{\theta})$). Using this assumption,

the Hamiltonian equations can be re-written as (Girolami & Calderhead, 2011):

$$\frac{d\boldsymbol{\theta}}{dt} = \mathbf{G}(\boldsymbol{\theta})^{-1}\mathbf{p}$$

$$\frac{dp_i}{dt} = -\frac{\partial J(\boldsymbol{\theta})}{\partial \theta_i} + \frac{1}{2}\mathbf{p}^T\mathbf{G}(\boldsymbol{\theta})^{-T}\frac{\partial \mathbf{G}(\boldsymbol{\theta})}{\partial \theta_i}\mathbf{G}(\boldsymbol{\theta})^{-1}\mathbf{p} \quad (7)$$

$$-\frac{1}{2}\text{tr}\left\{\frac{\partial \mathbf{G}(\boldsymbol{\theta})}{\partial \theta_i}\mathbf{G}(\boldsymbol{\theta})^{-1}\right\}$$

Moreover, the acceptance rate can be calculated as follows (Girolami & Calderhead, 2011):

$$\alpha = J(\boldsymbol{\theta}_k) + \frac{1}{2}\left\|\mathbf{G}(\boldsymbol{\theta}_k)^{-1}\mathbf{p}_k\right\|^2 + \frac{1}{2}\log\{|\mathbf{G}(\boldsymbol{\theta}_k)|\}$$

$$-J(\boldsymbol{\theta}_{k+1}) - \frac{1}{2}\left\|\mathbf{G}(\boldsymbol{\theta}_{k+1})^{-1}\mathbf{p}_{k+1}\right\|^2$$

$$-\frac{1}{2}\log\{|\mathbf{G}(\boldsymbol{\theta}_{k+1})|\} \quad (8)$$

When the Hamiltonian equations are not separable, the leapfrog algorithm in equation (5) is not applicable and the general form of leapfrog can be written as (Girolami & Calderhead, 2011):

$$\mathbf{p}_{k+1/2} = \mathbf{p}_k - \frac{\Delta t}{2}\nabla_\theta \mathcal{H}\{\mathbf{p}_{k+1/2}, \boldsymbol{\theta}_k\}$$

$$\boldsymbol{\theta}_{k+1} = \boldsymbol{\theta}_k + \frac{\Delta t}{2}\left[\nabla_p \mathcal{H}\{\mathbf{p}_{k+1/2}, \boldsymbol{\theta}_k\} + \nabla_p \mathcal{H}\{\mathbf{p}_{k+1/2}, \boldsymbol{\theta}_{k+1}\}\right]$$

$$\mathbf{p}_{k+1} = \mathbf{p}_{k+1/2} - \frac{\Delta t}{2}\nabla_\theta \mathcal{H}\{\mathbf{p}_{k+1/2}, \boldsymbol{\theta}_{k+1}\} \quad (9)$$

Many studies propose implicit or explicit methods for solving equation (9). However, to do so we need to calculate the derivative of $\mathbf{G}(\boldsymbol{\theta})$ w.r.t $\theta_i$, which means that we need to calculate $d$ different matrices in each iteration, that is computationally expensive.

In this study, we calculate the local Hessian at the start of each step and consider it constant during the step to avoid calculation of third derivative in each iteration, the updating formulation of method will be similar to the HMC method with a dynamic mass matrix that is calculated at the start of each step.

### 3.2 *Hessian-informed HMC method*

Some studies investigate using the Hessian-informed mass matrix to improve the performance of the HMC method (Bui-Thanh & Girolami, 2014). In (Karimi et al., 2021) we showed that using Hessian at the MAP point as the mass matrix can significantly accelerate the convergence speed of the HMC method and is computationally efficient, especially for nonlinear high-dimensional problems.

Consequently, the acceptance rate for this method can be calculated as:

$$\alpha = J(\boldsymbol{\theta}_k) + \frac{1}{2}\left\|\mathbf{H}_{\text{MAP}}^{-1}\mathbf{p}_k\right\|^2 - J(\boldsymbol{\theta}_{k+1})$$

$$-\frac{1}{2}\left\|\mathbf{H}_{\text{MAP}}^{-1}\mathbf{p}_{k+1}\right\|^2 \quad (10)$$

## 4 NUMERICAL RESULTS

In this section, we discuss a problem related to inferring the permeability field of a soil layer from sparse and noisy pressure data which are sensor measurements (Karimi et al., 2021). This is a large-scale inverse problem governed by a coupled partial differential equation (PDE), which is addressed by the Bayesian inference framework, and discussed in detail by Karimi et al., 2021.

In this example, we consider similar geometrical domain a $8{,}000 \times 4{,}000$ m layer, and $\boldsymbol{\theta}$ is the uncertain permeability field. We assumed a log-normal distribution, where the potential function is defined as follows:

$$J(\boldsymbol{\theta}) = \frac{1}{2}\left\|\boldsymbol{\Sigma}^{-1/2}(\log(\boldsymbol{\theta}) - \mathbf{m})\right\|^2 - \log(\prod_{i=1}^d \theta_i^{-1})$$

$$-\frac{1}{2}\log\left(\left|\boldsymbol{\Sigma}^{-1}\right|\right) \quad (11)$$

where $\boldsymbol{\Sigma}^{-1}$ and $\mathbf{m}$ are the covariance matrix and mean vector of the log of $\boldsymbol{\theta}$, respectively. The MAP point is:

$$\boldsymbol{\theta}_{\text{MAP}} = \exp(\mathbf{m} - \boldsymbol{\Sigma}\mathbf{1}) \quad (12)$$

In this example we consider $d = 936$, and the $\Sigma$ and $\mathbf{m}$ inputs are provided in the GitHub link reported at the end of this paper. Figure (1) shows the MAP point in the rectangular domain.

We generate 25,000 samples from the target distributions using four different methods, MH-MCMC, HMC, RMHMC using local Hessian at the start of each step (H(local)-HMC), and Hessian-informed HMC using Hessian at the MAP point (H(MAP)-HMC). The fixed step size for MH-MCMC, HMC, H(local)-HCM, and H(MAP)-HMC are assumed as $\Delta t = 0.01$, $0.15$, $0.3$, and $0.3$, respectively.

Figure 2 shows the autocorrelation of the spatial average value of random variable $\boldsymbol{\theta}$ versus lag, to compare the performance of different sampling methods. As the results show, H(local)-HMC and H(MAP)-HMC have similar performances. Moreover, the autocorrelation function for H(local)-HMC and that for H(MAP)-HMC decay to zero faster and show that these two algorithms explore the distribution significantly faster than MH-MCMC and HMC algorithms.

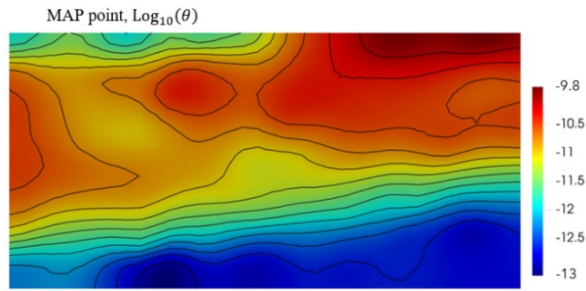

Figure 1. MAP point.

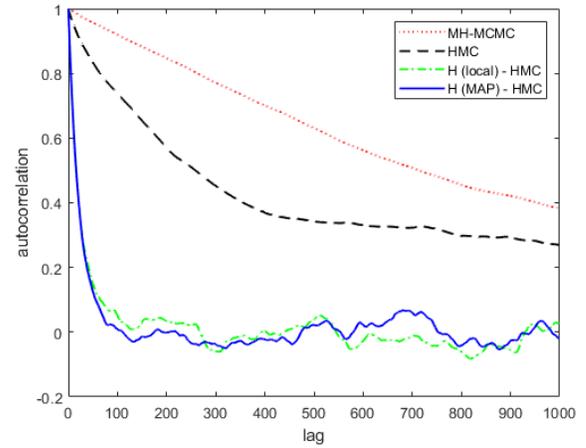

Figure 2. Autocorrelation vs. lag for different sampling methods, autocorrelation is calculated for spatial average value of $\boldsymbol{\theta}$.

Table 1 displays the summary of convergence analysis of these five sampling algorithms. Where "acce" is the acceptance rate of each algorithm, and $\tau$ is the correlation time which is defined as:

$$\tau = 1 + \sum_{t=1}^{\infty} \rho_t \qquad (13)$$

$\rho_t$ is defined as the correlation between two states in the chain with lag $t$. We have truncated the summation at its maximum value. Also, $N_{eff}$ is the effective sample number which is define as $N/\tau$. The results show an approximately similar acceptance rate for H(local)-HMC and H(MAP)-HMC algorithms. However, correlation time of H(local)-HMC method is smaller which makes the effective sample number higher.

Table 1. Summary of convergence analysis.

| Method | acce | $\tau$ | $N_{eff}$ |
|---|---|---|---|
| MH-MCMC | 0.8 | 2276.7 | 11 |
| HMC | 0.86 | 1868.5 | 13 |
| H(MAP)-HMC | 0.92 | 176.77 | 141 |
| H(local)-HMC | 0.91 | 123.41 | 202 |

Figure 3 shows the 95% credible interval for the variable $\theta$, along the dashed line after taking 10,000 samples. As it can be seen in figure 3, the results from H(MAP)-HMC algorithm matches well with the exact confidence intervals.

computational costs. We have showed that Hessian-informed HMC method using the nonlocal Hessian at the MAP point and Hessian-informed HMC method using the constant local Hessian at the start of each

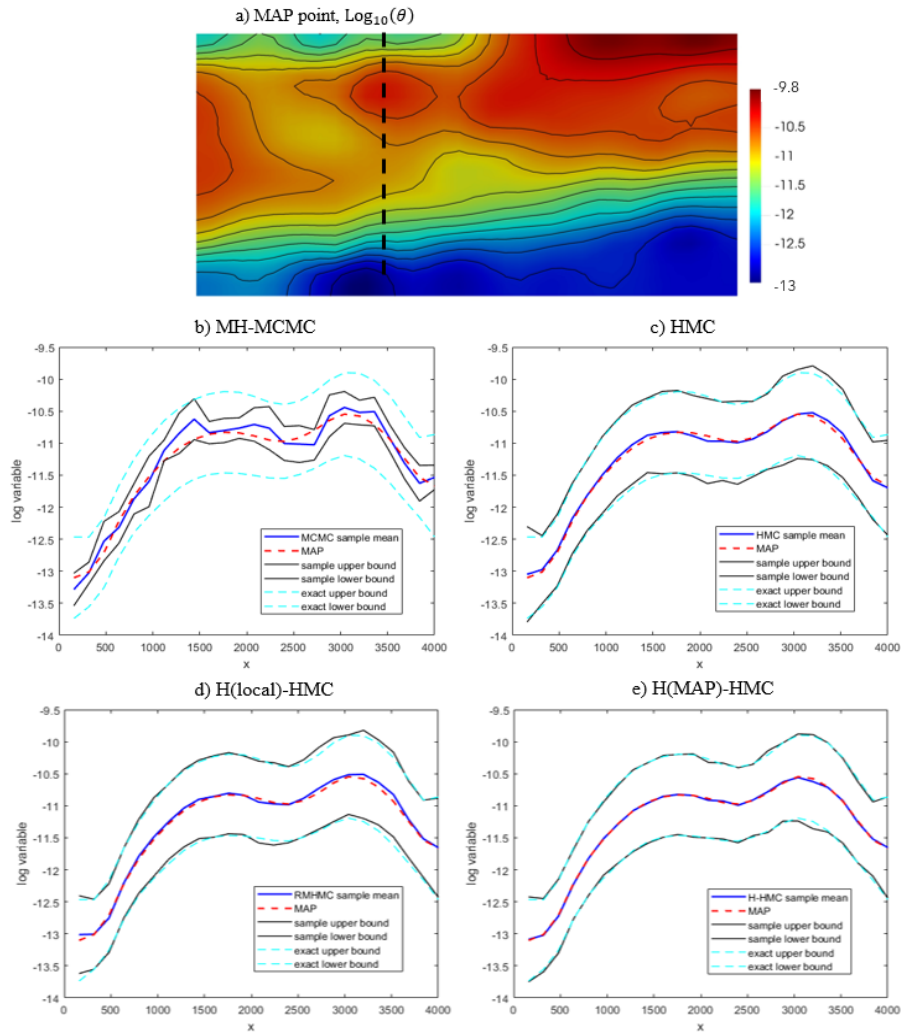

Figure 3. a) MAP point, b) 95% credible interval after 10,000 samples for MH-MCMC, c) HMC, d) H(local)-HMC, e) H(MAP)-HMC.

## 5 CONCLUSION

In this paper, we have investigated some Hessian-informed HMC algorithms. We have compared the performance of traditional MH-MCMC and HMC methods and the accelerated Hessian-informed algorithms. We have proposed using constant local second derivative information at the start of each iteration in the RMHMC algorithm to decrease the iteration can explore the high-dimensional distributions significantly faster than traditional sampling methods.

## 6 SOFTWARE AVALLABILITY

A version of the code developed for this work is available at:

https://github.com/minakari/Hamiltonian-Monte-Carlo

7 ACNOWLEDMENTS

We thank the National Science Foundation for support through XSEDE resources provided by Pittsburgh Supercomputing Center. Mina Karimi acknowledges financial support from the Scott Institute. Kaushik Dayal acknowledges financial support from NSF (CMMI MOMS 1635407, DMREF 1628994), ARO (MURI W911NF-19-1-0245), ONR (N00014-18-1-2528), BSF (2018183), and an appointment to the National Energy Technology Laboratory sponsored by the U.S. Department of Energy. Matteo Pozzi acknowledges financial support from NSF (CMMI 1638327). This work was funded (in part) by the Dowd Fellowship from the College of Engineering at Carnegie Mellon University; we thank Philip and Marsha Dowd for their financial support and encouragement.

8 PREFERENCES

Bui-Thanh, T., & Girolami, M. (2014). Solving large-scale PDE-constrained Bayesian inverse problems with Riemann manifold Hamiltonian Monte Carlo. *Inverse Problems*, *30*(11), 114014.

Chappell, M. A., Groves, A. R., Whitcher, B., & Woolrich, M. W. (2008). Variational Bayesian inference for a nonlinear forward model. *IEEE Transactions on Signal Processing*, *57*(1), 223-236.

Chen, W., Wang, Z., Broccardo, M., & Song, J. (2022). Riemannian Manifold Hamiltonian Monte Carlo based subset simulation for reliability analysis in non-Gaussian space. *Structural Safety*, *94*, 102134.

Cobb, A. D., Baydin, A. G., Markham, A., & Roberts, S. J. (2019). Introducing an explicit symplectic integration scheme for riemannian manifold hamiltonian monte carlo. *arXiv preprint arXiv:1910.06243*.

Girolami, M., & Calderhead, B. (2011). Riemann manifold langevin and hamiltonian monte carlo methods. *Journal of the Royal Statistical Society: Series B (Statistical Methodology)*, *73*(2), 123-214.

Girolami, M., Calderhead, B., & Chin, S. A. (2009). Riemannian manifold hamiltonian monte carlo. *arXiv preprint arXiv:0907.1100*.

Hairer, E., Hochbruck, M., Iserles, A., & Lubich, C. (2006). Geometric numerical integration. *Oberwolfach Reports*, *3*(1), 805-882.

Herbst, E. (2010). Gradient and Hessian-based MCMC for DSGE Models.

Karimi, M., Massoudi, M., Dayal, K., & Pozzi, M. (2021). High-dimensional nonlinear Bayesian inference of poroelastic field from pressure data.

Lee, Y. T., & Vempala, S. S. (2018, June). Convergence rate of Riemannian Hamiltonian Monte Carlo and faster polytope volume computation. In *Proceedings of the 50th Annual ACM SIGACT Symposium on Theory of Computing* (pp. 1115-1121).

Leimkuhler, B., & Reich, S. (2004). *Simulating hamiltonian dynamics* (No. 14). Cambridge university press.

Liu, Q., & Wang, D. (2016). Stein variational gradient descent: A general purpose bayesian inference algorithm. *arXiv preprint arXiv:1608.04471*.

Neal, R. M. (2011). MCMC using Hamiltonian dynamics. *Handbook of markov chain monte carlo*, *2*(11), 2.

Petra, N., Martin, J., Stadler, G., & Ghattas, O. (2014). A computational framework for infinite-dimensional Bayesian inverse problems, Part II: Stochastic Newton MCMC with application to ice sheet flow inverse problems. *SIAM Journal on Scientific Computing*, *36*(4), A1525-A1555.

Roberts, G. O., & Stramer, O. (2002). Langevin diffusions and Metropolis-Hastings algorithms. *Methodology and computing in applied probability*, *4*(4), 337-357.